\documentstyle[12pt]{article}

\def\Mpc{\,{\rm Mpc}}

\def\eV{{\,\rm eV}}

\def\cmm2{{\,\rm cm^{-2}}}
\def\cm2{{\,{\rm cm}^2}}
\def\cmm3{{\,{\rm cm}^{-3}}}
\def\gcmm3{{\,{\rm g\,cm^{-3}}}}
\def\kms{\,{\rm km\,s^{-1}}}

\def\la{\mathrel{\mathpalette\fun <}}
\def\ga{\mathrel{\mathpalette\fun >}}
\def\fun#1#2{\lower3.6pt\vbox{\baselineskip0pt\lineskip.9pt
  \ialign{$\mathsurround=0pt#1\hfil##\hfil$\crcr#2\crcr\sim\crcr}}}

\begin{document}
\baselineskip=14pt
\pagestyle{empty}
\begin{center}

\vspace{.2in}
{\Large \bf THE MEANING OF EROS/MACHO} \\

\vspace{.2in}
Michael S. Turner\\

{\it Departments of Physics and of Astronomy \& Astrophysics\\
Enrico Fermi Institute, The University of Chicago, Chicago, IL~~60637-1433}\\

\vspace{0.1in}

{\it Theoretical Astrophysics\\
Fermi National Accelerator Laboratory, Batavia, IL~~60510-0500}\\

\end{center}

\vspace{.3in}

\centerline{\bf ABSTRACT}
\bigskip

\noindent  Most of the mass density in the Universe---and in
the halo of our own galaxy---exists in the form of dark matter.  Overall,
the contribution of luminous matter (in stars)
to the mass density of the Universe
is less than 1\%; primordial nucleosynthesis indicates that
baryons contribute between 1\% and 10\% of
the critical density ($0.01h^{-2}\la \Omega_B\la 0.02h^{-2}$;
$h=$ the Hubble constant in units of $100\kms\Mpc^{-1}$);
and other evidence indicates that
the total mass density is at least 10\% of critical density,
and likely much greater.  If the universal density is as low as
10\% of the critical density there may be but one
kind of dark matter.  More likely, the universal
density is greater than 10\%, and
there are two kinds of dark matter, and thus two
dark matter problems:  In what form does the
baryonic dark matter exist? and In what form does the nonbaryonic
dark matter exist?  The MACHO and EROS collaborations have
presented evidence for the microlensing of stars in
the LMC by $10^{-1\pm 1}M_\odot$
dark objects in the halo of our own galaxy and may well
have solved {\it one} of the dark matter puzzles by
identifying the form of the baryonic dark matter.
It is too early to make precise statements about the fraction
of the mass density in the halo of our galaxy contributed by
lensing objects ($= f_m$), though the EROS/MACHO data
suggest that $f_m$ is probably 0.1 or larger.
Taking our galaxy to be typical and taking account
a fraction $f_m$ of the mass in the portion
of the halo that contributes most significantly to microlensing
(within 20\,kpc of the galactic center), I estimate that
lensing objects contribute a fraction $0.008f_m/h$ of the critical
density, and clearly cannot account for the bulk of
the dark matter if $\Omega_0\gg 0.1$ (even discarding
the nucleosynthesis bound).  (If the
distribution of lensing objects extends throughout the galactic halo,
out to about 100\,kpc,
their contribution is larger by a factor of five).
If halo lensing objects contribute around 10\% of the mass
density in the halo, the EROS/MACHO results provide further
evidence for a nonbaryonic component to the halo dark matter,
consistent with the simplest prediction for the
ratio of cold dark matter (e.g., axions or neutralinos)
to baryons in the halo.  In any case, the EROS/MACHO results
in no way lessen the strong case that now exists
for the mass density of the Universe being significantly larger
than that which baryons can contribute; nor do they
affect significantly the prospects for directly detecting
particle dark matter in our vicinity.  Discovering the nature of
the baryonic dark matter should provide still further
impetus for solving the more weighty dark matter problem,
the nature of the nonbaryonic dark matter.

\newpage
\pagestyle{plain}
\setcounter{page}{1}
\newpage

\section{Introduction}
\subsection{The events}

Two collaborations searching for massive, dark objects in the halo
of our galaxy through gravitational microlensing reported
candidate events this week.  The French EROS collaboration
has monitored the brightnesses of 3 million stars in the
Large Magellanic Cloud (LMC) over
a three-year period and reported two events:  a star
that brightened by a factor of about 2.5 over a characteristic
time interval of about 54 days and a star that brightened
by a factor of about 3.3 over a time interval of about 60 days \cite{eros}.
(EROS measure the brightnesses of stars
on $5^\circ \times 5^\circ$ Schmidt plates taken with two filters,
bleu and rouge, with a machine
known as ``MAMA.'')  The American-Australian MACHO collaboration
has monitored the brightnesses of about 1.8 million stars in the
LMC for one year and reported one event:  a star that
brightened by about a factor of about 6.8 over a characteristic
time interval of about 34 days \cite{macho}.\footnote{The definitions
of event duration used by MACHO and EROS differ by about a factor
of two; I have used the MACHO definition for all three events.}
(MACHO use two CCD cameras and a dichroic beamsplitter
to measure the brightnesses of stars
in two colors, red and blue, and observed each star about 250 times
during the year period.)

\subsection{Microlensing}

The basic idea of detecting massive, nonluminous objects
in the halo by microlensing was suggested by Paczynski \cite{pac};
the discussion that follows is based upon the very nice
paper by Griest \cite{griest}.  The actual deflection
of light from a star in the LMC (distance 50 kpc) by an
object in our halo is far too small to measure, of order $\delta\phi
\sim 2\times 10^{-4} (m/M_\odot)^{1/2}\,$arcseconds, where
$m$ is the mass of the halo object and the impact parameter
(distance between the lensing object and the line of sight
to the star) is
taken to be the Einstein radius (see below).  However, due to
gravitational focussing
the total brightness of the two unresolved images is greater
than that of the unlensed star, by a factor equal to 1.34
for an impact parameter equal to the Einstein radius (and
larger for smaller impact parameters).
Since the a priori brightness of a given star is not
known, Paczynski's idea was to look for time variation
of the brightness of the star due to the motion of the lensing object
across the line of sight.

The brightness amplification $A$ depends upon the impact
parameter $l$ in units of the Einstein radius $R_E$:
\begin{equation}
A = {u^2 +2 \over u(u^2+4)^{1/2}},
\end{equation}
where $l = u R_E$ and
\begin{equation}
R_E = \sqrt{4Gmx(L-x) \over c^2 L}.
\end{equation}
Here $x$ is the distance to the lensing object and
$L$ is the distance to the star.
A plot of $A(u)$ is shown in Fig.~1 (Fig.~2 of Ref.~\cite{griest});
large $A$ corresponds to small $u$.

Because the lensing object is moving (the {\it rms} halo
velocity is around $300\kms$) the impact parameter
changes with time, and hence the amplification does
too (in principle, one also must take into account the
velocity of the star and the solar system, though it is
a small correction \cite{griest}).  The time profile
of the amplification, or light curve $A(t)$,
depends upon $u(t)$ alone.

To be specific, the impact parameter $u$ is given by
\begin{equation}
u(t)^2 = u_{\rm min}^2 + (v_Tt/R_E)^2,
\end{equation}
where $u_{\rm min}$ is the minimum value of the impact
parameter, $v_T$ is the velocity of the lensing
transverse to the light of sight, and the epoch of maximum amplification
defines time zero.  Note that the light curves are
a two-parameter family of curves.  By the measuring the light curve
for a microlensing event one can infer $u_{\rm min}$ [$=u(A_{\rm max})$] and
$R_E/v_T$.  Recall, $R_E \sim GmL/c^2$, which implies
that the event duration $t\sim \sqrt{GmL}/c v_T
\sim 100\,{\rm da}\sqrt{m/M_\odot}$ (more later).

EROS define the duration of the event to be the characteristic
time $t_{\rm EROS}=R_E/v_T$, which corresponds to
impact parameter $u(\pm t_{\rm EROS}) =\sqrt{1+u_{\rm min}^2}$; MACHO
define the duration of the event to be the total time
the amplification is above threshold for detection
of amplification, $t_{\rm MACHO}
= 2 \sqrt{u_T^2-u_{\rm min}^2} R_E/v_T$.  For
maximum amplification $A_{\rm max} \gg 1$ and
a threshold amplification $A_T\sim 1.34$ ($u_T\sim 1$),
which seems to apply to the data of both collaborations,
$t_{\rm MACHO} \simeq 2t_{\rm EROS}$.

Now the formulas relating event rate to the
abundance of halo objects; to begin, assume that the lensing
objects have mass $m$, are distributed like the halo material,
and contribute a fraction $f_m$ of the halo density,
\begin{equation}\label{eq:halo}
\rho_m = f_ m {r_0^2+a^2\over r^2+a^2}\,\rho_{\rm local}.
\end{equation}
Here $r=$ distance from the center
of galaxy, $a=5\pm 5\,$kpc is the core radius of halo,
$r_0\simeq 8.5\,$kpc is the distance of the Solar System
from the galactic center, and $\rho_{\rm local} \simeq
5\times 10^{-25}\gcmm3$ is the local halo density.
(The mass density in the halo is determined by measurements
of the rotation curve of our galaxy and the amount and
distribution of light in the galaxy; the full extent of
the halo---and its total mass---are not known, though
there is evidence
that the halo extends at least as far as the LMC and probably
out to 100\,kpc \cite{tremaine}.)
Further, let us assume that the threshold for detection
is amplification $A_T$, corresponding to impact parameter
$u_T$, and that the efficiency for the detection of a
microlensing event above the amplification threshold
is $\varepsilon$.  (The latter is clearly an oversimplification as
$\varepsilon$ will be a function of $A$, the duration
of the event, and perhaps other things.)

The probability that a given star in the LMC is being microlensed
with amplification greater than $A_T$ by a halo object is referred to
as the optical depth for microlensing,
and was calculated by Griest \cite{griest}, cf. Eq.~(4),
\begin{equation}
\tau \simeq 4\times 10^{-7}f_m u_T^2;
\end{equation}
where the halo is assumed to extend out at least as far as the LMC,
the core radius is taken to be 5\,kpc, $u_T$ is the
impact parameter corresponding to $A_T$
($u_T=1$ for $A_T =1.34$).  The value
of $\tau$ is not very sensitive to the first two assumptions.

Another interesting quantity is $\tau (y)$, the probability
that a star in the LMC is microlensed by a halo object
between the galactic center and distance $y$ from it; $\tau (y)$
is shown in Fig.~2.  (Said another way, $\tau (y)$ is the
probability for microlensing under the assumption that the halo
only extends out to radius $y$.)  Figure 2 illustrates that most
of the optical depth for microlensing is due to halo
objects within $10-20\,$kpc or so
of the center of the galaxy.\footnote{The reason is simple:
the Einstein radius, which the sets the radius of the
tube within which halo objects lead to amplification
above threshold, is proportional to
the product of the distance to the lens and the distance
from the lens to the LMC, and further, $d\tau /dy\propto
\rho_m/m$, which falls off as $r^{-2}$.}
{\it That is, microlensing only
probes the inner portion of the halo, and further, not
much of the lensing probability is due to objects
near the solar neighborhood.}  This is an important fact
to keep in mind when interpreting the EROS/MACHO events.

What makes microlensing observable is the time variation
of the amplification.  So even more important than
the lensing probability is the rate of microlensing events
(roughly $\tau$ divided by the typical event
duration); integrating
over the distribution of velocities and
positions of halo lensing objects Griest obtains the microlensing rate
(per star observed) with amplification greater than $A_T$, cf.~Eqs.~(7-14),
\begin{equation}
\Gamma \simeq 1.7\times 10^{-6} f_m
u_T/(m/M_\odot )^{1/2}\,{\rm yr}^{-1}.
\end{equation}
Note, $\Gamma$, unlike $\tau$, depends upon the mass
of the halo objects:  for fixed $f_m$, the smaller the
mass, the higher the event rate.

The duration and amplification of a microlensing event depends upon the
impact parameter and the velocity, mass, and distance of the halo microlens.
However, the velocity and distance of the halo lensing object
can only be specified statistically, and thus the mass
of the lensing objects can only
be inferred from a measured light curve in a statistical way.
Griest has constructed the likelihood function
for the lensing mass, given the event duration and threshold
for detection.  The distribution is very broad, about a order of magnitude
in mass at full-width, half maximum (Fig.~9 in
Ref.~\cite{griest}).  The peak of that likelihood function occurs at a mass
\begin{equation}
\langle m \rangle \simeq 1.5M_\odot (t/100\,{\rm da})^2/u_T^2.
\end{equation}
Note that this mass depends upon the threshold of the experiment;
for EROS and MACHO $u_T\sim 1$.

The distribution of maximum amplification of microlensing
events is basically geometric:  The fraction of events
with maximum amplification greater than $A$ corresponds to
events with minimum impact parameter less than $u_{\rm min}
=u(A)$, which is simply equal to $u(A)/u_T$.  This fraction $\epsilon (A)$
can also be expressed in terms of $A$ and $A_T$,
\begin{equation}
\epsilon (A) = \sqrt {{(1-A^{-2})^{-1/2}-1\over
(1-A_T^{-2})^{-1/2} -1}},
\end{equation}
and is show in Fig.~3; cf.~Fig.~10 of Ref.~\cite{griest}.

To conclude this section; microlensing has many clear signatures:
it is achromatic; the light curve is defined uniquely
in terms of two parameters, maximum amplification and
duration; the distribution of event
amplifications is predicted; microlensing shows
no preference for the type of lensed star; and so on.
The most significant background for microlensing searches
are variable stars; already, MACHO has compiled the largest
catalogue of variable stars in the LMC.  Because of its
many signatures, microlensing should be a relatively
easy hypothesis to test; indeed, EROS and MACHO have
already mastered the difficult tasks of monitoring the
brightnesses of millions of stars and discriminating
against known types of variable stars.  In the next year or so
both collaborations should increase their data
sets by a factor of ten or so and thus should clarify the
few questions that have been raised about their
events (large $\chi^2$ for two of the three events
and large amplification for the MACHO event despite their
low threshold, cf. Fig.~3).  The case for (or possibly
against) the existence of halo lensing objects
should be firmly established soon.

\section{Overinterpreting the Events}

What can one learn from the very preliminary MACHO and EROS
results?  Definitely not as much as one would like!  But let's
risk going out on a limb with some shaky interpretations.

First, the number of events seen.  Based upon their
stated exposures the number of expected events is
\begin{eqnarray}
N_{\rm MACHO} & = & 9.4 f_m \varepsilon u_T /(m/0.1M_\odot )^{1/2},\\
N_{\rm EROS} & = & 47 f_m \varepsilon u_T /(m/0.1 M_\odot )^{1/2}
\end{eqnarray}
where of course the detection efficiencies and thresholds
for the two experiments are not necessarily the same.
In order to infer the
fraction of the mass density of the halo in lensing objects, $f_m$, one
must know not only the efficiencies, but also something
about the mass of the lensing objects (see below).\footnote{What
one wants to do, and what both collaborations are certainly
doing, is constructing the likelihood function for their events
as a function of all the relevant parameters (lens mass or
distribution of lens masses, spatial
density of lenses, and so on) and taking account of
acceptance, etc.  From this, the most likely values for
$m$ and $f_m$ and estimates for their variances can be inferred.}

MACHO has made no statement about $f_m$.  From the duration
of their event they infer a most likely mass of about
$0.1M_\odot$; from their paper and conversations with collaboration
members I conclude that both $u_T$ and $\varepsilon$ are of order
unity.  Climbing out to the very end of the limb,
this suggests that $f_m$ is about 0.1 or larger (with great uncertainty).

EROS state in their paper that ``the number of events
is consistent with the hypothesis that the halo of our
galaxy is made essentially of dark objects in the mass
range of a few $10^{-2}$ to a few $M_\odot$.''
Taking $m =0.1M_\odot$, this implies an
efficiency $\varepsilon u_T \sim 0.04$.  My interpretation
of a talk given by EROS member Luciano Moscoso is that,
based upon Monte Carlo simulations,
the detection efficiency for events with $u_{\rm min}\la
1$ is about 80\% and for events with $u_{\rm min}\ga 1$ it is much,
much smaller.  Their exposure seems to be much lower than
3 million stars times 3 years:  only about half the stars
were used and the time coverage of a typical star was
about 600 days.  Putting this together, I guess that $u_T\sim 1$
and $\varepsilon\sim 0.1$, suggesting that $f_m\sim 0.2$ or so.

Next, event duration; as mentioned earlier, the
likelihood function computed by Griest
peaks at a mass of $1.5M_\odot (t/100\,{\rm da})^2/u_T^2$.
For MACHO, this implies a lensing
mass of about $0.1M_\odot$, which is the central value stated
in their paper.  For EROS, the event durations are
about twice as long; this implies
a most likely mass of about $0.4M_\odot $.
While the estimates for the lensing mass seem to be
different, recall that the likelihood function
is an order of magnitude in mass at FWHM and that
the detection efficiencies will certainly depend
upon the event duration and can influence the likelihood function.

As noted above, both the event rate and event duration are
needed to infer the fraction of critical density
contributed by halo lensing objects.  Moreover,
some a priori information may help in this regard.
For example, stars made of hydrogen and helium more massive than
about $0.1M_\odot$ are nuclear burning; previous
unsuccessful searches for low-mass stars (M dwarfs) place
severe limits to $f_m$ for $m\ga 0.1M_\odot$, perhaps
even ruling out this possibility \cite{mdwarf,carr}.  Dark halo
objects heavier than around a solar mass would have to be
neutron stars, black holes, or white dwarfs, otherwise they would
be easily visible; however, this possibility too
is severely constrained by their contribution to the
heavy-element abundance in our galaxy \cite{carr}.  For these reasons
it has been argued very convincingly that the
mass of baryonic halo objects must be less than about
$0.1M_\odot$; note, this makes my previous shaky
estimates for $f_m$
shaky {\it upper limits} since $\Gamma \propto m^{-1/2}$
and $f_m\propto m^{1/2}$.

Let's go on and estimate the fraction of critical density
contributed by halo lensing objects, $\Omega_m$.
The strategy is as follows:  (1) construct the ratio
of mass in halo lenses
to light for the Milky Way, $M_m/L$; (2) assume that this ratio
is universal; (3) Divide this ratio by the critical mass to light
ratio to find $\Omega_m$.\footnote{Astronomers measure
the mass density of the Universe by multiplying the luminosity
density, ${\cal L} \simeq 2.4h\times 10^8 L_{B\odot}\Mpc^{-3}$,
times a characteristic mass to light ratio; the
mass to light ratio (in the $B_T$ system) that corresponds
to critical density is $1200h M_\odot /L_{B\odot}$.}

Much is known about the halo mass interior
to about 20\,kpc \cite{tremaine}, and recall, most of the microlensing
is due to objects within 20\,kpc of the galactic center.
The halo mass interior to 20\,kpc is about $2.3\times 10^{11}M_\odot$.
The luminosity of the Milky Way galaxy is about
$2.3\times 10^{10}L_{B\odot}$.  Assigning a fraction
$f_m$ of the halo mass within 20\,kpc of the galactic
center to lensing objects, I find
\begin{eqnarray}
{M_m\over L} & \simeq & 10f_m {M_\odot \over L_{B\odot}} \\
\Omega_m & \simeq & 8\times 10^{-3}f_m/h .
\end{eqnarray}

This is not much mass, though, to be sure, at least one of our assumptions
was very conservative.  Although
microlensing searches are most sensitive to halo lensing objects
within 20\,kpc of the galactic center, it would be surprising
if $f_m$ suddenly dropped to zero at this point.  If
we assume that halo lensing objects
populate the halo at the same fractional abundance
out to 100\,kpc, the known extent of the
halo, the previous estimate increases by a factor of five:
\begin{equation}
\Omega_m \simeq 0.04f_m/h,
\end{equation}
though still far from closure density.

To conclude this discussion, the MACHO and EROS collaborations
may well have solved {\it one} of the dark matter riddles, namely,
what form the dark baryons take.  If so, their searches were spectacularly
successful.  However, the weightier dark matter problem is
the nature of the nonbaryonic dark matter, as evidence
indicates that $\Omega_0$ exceeds 0.1, the
maximum contribution permitted for baryons.
(Nucleosynthesis bound aside,
the contribution of halo lensing objects to the universal
density cannot account for the total mass
density if $\Omega_0$ is significantly greater than 0.1.)

\section{Implications for Particle Dark Matter}

There is mounting evidence that $\Omega_0$ is significantly
greater than 0.1, the maximum contribution of baryons,
and perhaps close to unity.
The evidence that $\Omega_0$ is close to unity
includes several studies relating the peculiar velocities of
galaxies within a few $100\Mpc$ of the Milky Way (and the
Milky Way itself) to the distribution of galaxies determined
from red-shift surveys based upon the IRAS Point
Source Catalogue\footnote{The peculiar velocities of
galaxies arise due to gravitational forces resulting from
the inhomogeneous distribution of matter and depend upon
the distribution of galaxies and the mean mass density;
by determining the distribution of galaxies one can infer
the mean mass density; see e.g., Refs.~\cite{iras}.}
and the fact that the only models for the formation of structure in the
Universe that are consistent with measurements of
Cosmic Background Radiation (CBR) anisotropy
(e.g., COBE and experiments on smaller
angular scales) and large-scale structure
require nonbaryonic dark matter.
A number of measurements of $\Omega_0$, e.g., cluster virial
masses and the infall of nearby galaxies into the
Virgo Cluster, strongly suggest that $\Omega_0$
is greater than 0.1, though perhaps not as large as unity.
And finally, there are theoretical reasons in favor
of $\Omega_0 =1$ and nonbaryonic dark matter:  a flat Universe is
an unambiguous prediction of the inflationary paradigm,
and there are several particles whose postulated existence is motivated
by compelling particle-physics considerations and whose relic
abundance is close to the critical density (e.g., neutralino and axion)
\cite{pdm}.

In a flat, critical Universe particle dark matter is
obligatory.  For the discussion that follows I assume that
$\Omega_0 =1$ and that $\Omega_B \sim 0.05 - 0.10$.
The universal ratio of exotic dark matter to baryonic dark matter is then
\begin{equation}
{\Omega_X\over \Omega_B}= {1-\Omega_B\over \Omega_B}\simeq 10-20.
\end{equation}
Of greater interest for the interpretation of the MACHO/EROS events is
the local ratio of particle
dark matter to baryonic dark matter.  In order to fully
answer this question we must know how galaxies and
other structures formed.  In the simplest
circumstance, where only gravity is important and
the initial velocities of baryons and particle dark matter are
similar (or negligible), the answer is dictated by the
equivalence principle:  Baryons and particle dark matter
must follow the same trajectories so that the ratio remains constant
and equal to $10-20$.

When nongravitational forces become important then the
two forms of dark matter will
certainly become differentiated as particle dark
matter does not interact electromagnetically, the most important
nongravitational force that operates in the Universe.  In general, the
interactions of baryonic matter with itself (and the CBR)
allow baryons to ``cool'' (i.e., dissipate their energy)
and become more condensed, thereby decreasing the ratio of particle dark
matter to baryonic dark matter.

For hot dark matter, relic particles that move fast (e.g.,
$20\eV$ neutrinos), the velocity distribution of the particle dark matter
is very important:  Neutrinos move too fast to become trapped
in galaxies (at least until very recently) and so one expects the
local ratio of hot dark matter to baryonic matter to be much, much
lower than the universal ratio of $10-20$.

For cold dark matter (e.g., axions or neutralinos), relic
particles that move very slowly, the velocity distribution is
unimportant and one would expect that in objects whose formation
only involves gravity the local density would reflect the
universal value of $10-20$.  There is no doubt that the formation
of the disk of our galaxy involved dissipation (the disk-like
structure traces to the dissipation of essentially all the
energy not associated with angular momentum).  While the
halos of galaxies are not fully understood, they show no obvious
signs of dissipation having been involved in their formation.
Thus, the simplest
assumption is that the ratio of cold dark matter to baryonic
dark matter in the halo should reflect the
universal value of $10-20$.\footnote{The ratio of cold dark matter
to baryonic dark matter expected for the mixed dark matter model ($\Omega_B
\sim 0.05$, $\Omega_{\rm cold} \sim 0.65$, $\Omega_{\rm hot}\sim
0.3$) is only slightly smaller, around $7-13$.  This is one
of the models that best accounts for the observed large-scale
structure.}

The fraction of the halo mass density
in lensing objects is crucial to understanding the implications
of MACHO and EROS for particle dark matter.  Let us consider two
possibilities consistent with the data at hand:
$f_m\sim 0.1$ and $f_m\sim 0.5$.

Suppose $f_m$ turns out to be of order 0.1, as is suggested
by a naive overinterpretation of the data.  In this case
the MACHO/EROS results are consistent with the simplest
expectation for cold dark matter and inconsistent with
hot dark matter.  Moreover, the fact that 90\% of
the mass density in the halo remains unexplained would
provide even further impetus for searching
for cold particle dark matter in our own halo.

On the other hand, suppose that two or three years from now
EROS and MACHO convincingly establish that $f_m$ is
0.5.  This is consistent with
hot dark matter and inconsistent with the simplest expectation
for cold dark matter.  With regard to the latter I mention:

\begin{enumerate}

\item It could be that the inner part of the halo (which is
what is probed by MACHO and EROS) has a smaller ratio of
cold dark matter to baryons because the baryons underwent
some dissipation (after all they cannot have formed into
astrophysical objects without dissipation).  At present
there are no compelling scenarios for the formation of
these objects, so it is certainly possible that the baryons
in the inner halo have undergone some dissipation.

\item It could be that the universal ratio of cold dark
matter to baryons is only a few.  One of the models
that best accounts for the observed large-scale
structure is cold dark matter + cosmological
constant ($\Omega_\Lambda =0.8$; $\Omega_{\rm cold}\sim
0.15$; and $\Omega_B \sim 0.05$).  While it is certainly
not the best motivated model, if the
Hubble constant is close to $80\kms \Mpc^{-1}$, as many
now believe \cite{hogan}, a cosmological constant will be
mandatory to solve the cosmological age crisis.

\item A halo ratio of cold dark matter to baryonic dark matter
of order unity, rather than $10-20$, has only a minor impact
on experimental searches for the cold dark matter in our vicinity,
as these experiments already take into account a
factor of two uncertainty in the local density of halo
material.\footnote{Given the uncertainties about the halo
mass, even within 20\,kpc of the galactic center,
and the insensitivity of microlensing to the density of lensing objects
nearby, it seems unlikely that microlensing experiments
could ever establish with confidence that
lensing objects contribute more than about 50\% or so of the local
dark matter density.  Moreover, because there is nothing to prevent
cold dark matter from falling into the halo of our galaxy,
even if baryonic matter has been concentrated by dissipation,
it is difficult to imagine that the local density of cold dark matter
could be significantly less than the local halo density.}

\end{enumerate}

\section{Concluding Remarks}

Luminous matter (in stars) contributes less than 1\% of the critical
density, and primordial nucleosynthesis makes a strong
case that ordinary matter contributes between 1\% and 10\%
of the critical density, establishing the existence of the
first dark matter puzzle, the form of
the baryonic matter.  There is compelling
evidence that the mass density of the Universe is much greater
than 10\% of the critical density and perhaps close to
the critical density,
establishing the existence of a second dark matter puzzle,
the form of the nonbaryonic dark matter.

The EROS/MACHO collaborations may well have solved the first
dark matter problem by providing evidence for dark baryons
in the form of astrophysical mass objects in the halo of
our own galaxy.\footnote{Dark baryons may well exist in
more than one form.  While only about 5\% of galaxies
are found in rich clusters of galaxies, in clusters
most of the baryons exist in the form of hot, x-ray emitting
gas (that is dark only in the context of ``visible radiation.''}
However, their discovery, as important
as it is, has little to say about the second dark matter
problem:  The mass density contributed by halo lensing objects
cannot significantly exceed 10\% of the critical density
(even discarding
the constraint to $\Omega_B$ based upon primordial nucleosynthesis).

A critical Universe is well motivated, supported by mounting
evidence, and must involve particle dark matter.
The detection of dark stars in the halo of our own galaxy
changes none of this; nor does it lessen significantly the prospects for
the direct detection of the cold dark matter in our vicinity
(e.g., in laboratory experiments designed to detect
halo axions or neutralinos).  The scientific stakes in testing the
particle dark matter hypothesis are extremely high:  identifying
the primary constituent of the Universe, discovering a
new particle of nature, and probing the earliest moments
of the Universe.  The pioneering searches
for particle dark matter \cite{search} must continue at full speed; if
anything, solving part of the dark matter puzzle should
provide still further impetus for solving the rest of the puzzle!

\bigskip\bigskip

It is a pleasure to thank Evalyn Gates and Rene Ong for
valuable comments.  This work was supported in part
by the Department of Energy
(at Chicago) and by the NASA through grant NAGW-2381 (at Fermilab).

\vfill\eject
\section{Figure Captions}
\bigskip

\noindent{\bf Figure 1:}  Impact parameter $u$ (in units
of the Einstein radius, $u=l/R_E$) as a function of the
amplification.

\medskip
\noindent{\bf Figure 2:}  The optical depth for gravitational
lensing due to halo objects within a distance $y$ of the
galactic center, $\tau (y)$, as a function of $y$.
Note that most of the optical depth for microlensing is
due to objects within $10-20\,$kpc of the galactic center.
Thus, microlensing experiments only probe this part of the halo
and are insensitive to the extent of the halo as well
as the local halo density.

\medskip
\noindent{\bf Figure 3:}  The fraction of microlensing events
expected with amplification greater than $A$ for
amplification thresholds of $A_T= 1.1, 1.34, 1.5, 2, 3,$ and $5$.


\begin{thebibliography}  {pac}

\bibitem{eros}  E.~Aubourg et al. (EROS Collaboration),
{\it Nature}, in press (1993).

\bibitem{macho} C.~Alcock et al. (MACHO Collaboration),
{\it Nature}, in press (1993).

\bibitem{pac} B.~Paczynski, {\it Astrophys. J.} {\bf 304}, 1 (1986).

\bibitem{griest} K.~Griest, {\it Astrophys. J.} {\bf 366}, 412 (1991).

\bibitem{tremaine} M.~Fich and S.~Tremaine, {\it Ann. Rev.
Astron. Astrophys.} {\bf 29}, 409 (1991).

\bibitem{mdwarf} See e.g., G.~Gilmore and P.~Hewett, {\it Nature}
{\bf 306}, 669 (1983); D.~Richstone
et al., {\it Astrophys. J.} {\bf 388}, 354 (1992).

\bibitem{carr} See e.g., B.J.~Carr, {\it Nucl. Phys. B} {\bf 252}, 81 (1985);
or D.J.~Hegyi and K.A.~Olive, {\it Astrophys. J.} {\bf 303}, 56 (1986).

\bibitem{iras} M.~Rowan-Robinson et al.,
{\it Mon. Not. R. astr. Soc.} {\bf 247}, 1 (1990);
N.~Kaiser et al., {\it ibid} {\bf 252}, 1 (1991);
M.~Strauss et al., {\it Astrophys. J.} {\bf 385}, 444 (1992);
E.~Bertschinger and A.~Dekel, {\it Astrophys. J.}
{\bf 336}, L5 (1989); A.~Dekel et al., {\it Astrophys. J.}, in press
(1993).

\bibitem{hogan}  M.~Fukugita, C.J.~Hogan, and P.J.E.~Peebles,
{\it Nature}, in press (1993).

\bibitem{pdm} See e.g., M.S.~Turner, {\it Physica Scripta}
{\bf T36}, 167 (1991); or {\it Proc. NAS} {\bf 90}, 4827 (1993).

\bibitem{search}  See e.g., P.F.~Smith and J.D.~Lewin, {\it Phys.
Repts.} {\bf 187}, 203 (1990); D.O.~Caldwell, {\it Mod. Phys. Lett. A}
{\bf 5}, 1543 (1990); K.~van Bibber et al., in {\it Trends in
Astroparticle Physics}, edited by D.~Cline and R.D.~Peccei
(WSPC, Singapore, 1992), p.~154.

\end{thebibliography}
\end{document}